# Interplay between itinerant and localized states in CaMn$_{1-x}$Ru$_x$O$_3$ ($x \leq 0.5$) manganites

V. Markovich,[a*] M. Auslender,[b] I. Fita,[c,d] R. Puzniak,[c] C. Martin,[e] A. Wisniewski,[c] A. Maignan,[e] B. Raveau,[e] and G. Gorodetsky[a]

[a]*Department of Physics, Ben-Gurion University of the Negev, P.O. Box 653, 84105 Beer-Sheva, Israel*

[b]*Department of Electrical and Computer Engineering, Ben-Gurion University of the Negev, P.O. Box 653, 84105 Beer-Sheva, Israel*

[c]*Institute of Physics, Polish Academy of Sciences, Aleja Lotnikow 32/46, PL-02-668 Warsaw, Poland*

[d]*Donetsk Institute for Physics and Technology, NAS, R. Luxemburg str. 72, 83114 Donetsk, Ukraine*

[e]*Laboratoire CRISMAT, UMR 6508, ISMRA, 6 Boulevard du Marechal Juin, 14050 Caen Cedex, France*

## Abstract

Magnetic properties of polycrystalline CaMn$_{1-x}$Ru$_x$O$_3$ ($x = 0 - 0.5$) samples were investigated in the temperature range 4.2 – 250 K, under external magnetic field up to 15 kOe and under hydrostatic pressure up to 12 kbar. Transport properties of the samples with $x = 0.1, 0.2, 0.4, 0.5$ were also investigated under pressure up to 10 kbar. For $x$ up to 0.4, the pressure was found to suppress ferromagnetic correlations and to increase the resistivity, while for $x = 0.5$ to act in the opposite way. While long ferromagnetic order is completely suppressed, in small clusters ferromagnetic correlations probably survive under pressure, as was revealed for CaMn$_{0.9}$Ru$_{0.1}$O$_3$. The pressure effect on the magnetic interactions and on the volume of ferromagnetic phase was found to depend strongly on the Ru-content, and absolute value of the pressure coefficient of spontaneous magnetization was found to decrease practically linearly with increasing $x$ in the range $0.1 < x < 0.5$. The experimental data are discussed in the frame of proposed energy-level diagram, which includes magneto-impurity states at low and moderate Ru-content and mixed-valence states of Ru presented by a strongly-correlated $t_{2g}$–like band at heavy Ru-doping. An impact of disorder introduced by Ru-doping on the energy diagram and on derived magnetic interactions is discussed. Predictions of the model regarding the pressure effects on conductivity and temperature scales characteristic for magnetic interactions are in reasonable agreement with experiment.

## I. INTRODUCTION

It is a matter of consensus that mechanism ruling ferromagnetism in the perovskite-structure manganites is double exchange (DE) mediated by hopping of electrons between manganese ions, which facilitates both ferromagnetic (FM) order and electrical conductance, thereby resulting in a ferromagnetic metallic (FMM) phase. DE was proved also to account for the electronic/magnetic phase separation (PS).[1] The presence of states, at which excess carriers remain localized close to impurity or manganese ion, may energetically favor a super exchange (SE) like interactions, which may yield an FM insulating (FMI) or antiferromagnetic (AFM) phases.[2] Coexistence of FMM domains with AFM regions was found in many manganites.[1,2] Magnetic and electric fields, and applied pressure as well, may vary the subtle balance between coexisting phases.[1,2]

The substitution of Ca in $CaMnO_3$, being G-type AFM insulator with Néel temperature $T_N \sim$ 120 K, by ions with valence larger than +2, e.g. La and Sm, which dope electrons, results in complicated structural changes. In $Ca_{1-x}La_xMnO_3$ and $Ca_{1-x}Sm_xMnO_3$ the crystal structure transitions and accompanying magnetic transitions:

P$nma$ – G-AFM → P$nma$ – G-AFM/FM + P$2_1/m$ – C-AFM (mixed phase) → P$2_1/m$ – C-AFM

occur, when $x$ increases from 0 to 0.2,[3-7] because of capturing the doped electrons by $Mn^{4+}$ ions in larger and larger portion. Due to emergence of long-living $Mn^{3+}$ ions, Jahn-Teller (JT) effect, charge and orbital ordering (CO and OO) occur, making these compounds strong insulators. Unlike that, substitution of Mn by ions with valence larger than +4, e.g. Ru and Mo, which also dope electron may induce a FM state and electric conduction.[8-10] In $CaMn_{1-x}Ru_xO_3$, parent crystallographic structure remains undistorted with increasing $x$ up to $x = 0.8$.[8-11] The magnetic structure, however, evolves drastically in comparison to that of $CaMnO_3$.[9,11] Substituting Mn in $Ca_{0.8}Sm_{0.2}MnO_3$ by a small amount of Ru is sufficient to remove monoclinic distortions, to impede CO and to form quasi metallic FM phase.[11-15] The pressure effect on the magnetic and transport properties of $Ca_{0.8}Sm_{0.2}Mn_{1-x}Ru_xO_3$ has been shown to depend crucially on $x$.[16,17] On the contrary, the Mo-substitution, which yields initially FMM state, drives $CaMn_{1-x}Mo_xO_3$ back to an AFM insulator (though being of C-type) state yet at $x \approx 0.06$. [18] The monoclinic distortions result from cooperative JT effect, which indicates the presence of stable $Mn^{3+}$ ions. Hence, the above-noted properties demonstrate unambiguously that, to the best of nowadays knowledge, Ru is unique electron dopant which prevents appearance of the JT active $Mn^{3+}$ ions. [15] On the other hand, our recent study [19] of low-doped $CaMn_{1-x}Ru_xO_3$ ($0 \leq x \leq 0.15$) under pressure has demonstrated a collapse of long-range FM order at $x = 0.1$ under pressure of about 10 kbar. These intriguing peculiarities motivated the present work.

In this paper we present detailed report on further experimental results regarding the effect of hydrostatic pressure on the magnetic and conductive properties of $CaMn_{1-x}Ru_xO_3$ in the range of $0 \leq x \leq 0.5$, which is much wider than explored before (for a narrower range, some data have been in short communicated recently [20]). We use the results of the present and previous [8,11,19-21] studies in a conjunction with knowledge of electronic structure to suggest a theoretical model. The emerging energy-level diagram qualitatively explains the dependence of the Curie and Neel temperatures and spontaneous magnetization on pressure and of the resistivity on pressure and temperature, in both diluted and concentrated $CaMn_{1-x}Ru_xO_3$, though the quantitative description remains a challenge at the present stage. Nevertheless, the proposed model may serve as a prototype for future first-principle modeling of the manganite-ruthenate oxides.

## II. EXPERIMENTAL

### A. Samples and experimental setups

Studies were carried out on polycrystalline samples of $CaMn_{1-x}Ru_xO_3$ ($0 \leq x \leq 0.5$), prepared by a standard ceramic route, starting from the stoichiometric ratios of CaO, $MnO_2$ and $RuO_2$, with intermediate crushing and heating. [9] All samples were found to be compatible with the orthorhombic perovskite structure of P*nma* space group within the temperature interval $1.5 \leq T \leq 300$ K. [8-10] Methods of the samples' structural characterization (XRD, EDS etc.), as well as their magnetic and transport properties at ambient pressure have been reported in Refs. 8-9. According to dc-magnetic, transport, neutron-diffraction, ac-susceptibility and EMR studies [8-12] the samples with $x \geq 0.1$ are in a mixed G-AFM – FM phases' ground state.

We performed magnetic measurements under hydrostatic pressure by the use of a PAR Model 4500 vibrating sample magnetometer in the temperature range 4.2 – 250 K and in the magnetic fields up to 15 kOe. In this method a miniature container of CuBe with an inside diameter of 1.4 mm was used as a pressure cell. [16] The pressure at low temperatures was determined according to the known pressure dependence of the superconducting transition temperature of pure tin, placed near the investigated sample. Measurements of the resistivity ρ under pressure ($P$) up to $P = 10$ kbar and at temperatures $77 < T < 270$ K were carried out in another CuBe pressure cell with 6 mm inside diameter. In this case, we measured the temperature by a Cernox Resistance Thermometer, attached to the CuBe cell, while a manganin gauge [17] monitored the pressure. For customary four-point resistance measurements, we employed evaporated gold strips with a separation of about 0.3 mm between the voltage contacts. As a pressure-transmitting medium in the both above noted pressure cells, we utilized a mixture of mineral oil and kerosene.

### B. Pressure effects on magnetic properties and conductivity

Figures 1–4 depict the results of the magnetic measurements as the magnetization $M(T)$ curves at a given magnetic field ($H$), and $M$ vs. cycling $H$ (hysteretic) curves at $T = 5$ K, under ambient and various pressures. For the sample with $x = 0.5$ the $M(H)$ curves at different temperatures under ambient pressure are presented as well. We display $T$ dependences of field-cooled magnetization ($M_{FC}$) and zero field-cooled one ($M_{ZFC}$), which differ notably. For $x \geq 0.2$, the $M(T)$ curves have similar shape. In particular, they have high-temperature inflection points which are close for both $M_{FC}$ and $M_{ZFC}$ at given $x$. A $M_{ZFC}(T)$ curve has bell-like shape with a weak low-temperature local maximum, while on a respective $M_{FC}(T)$ curve the maximum becomes much less pronounced, if not smeared. Thus the features inherent for FM and AFM materials seem comprised in each of the measured $M(T)$ curves (the case of $x = 0.1$ is an exception [19]). By that reason, we use notation $T_C$ and $T_N$ for the above inflection and maximum point, respectively. Below we present a detailed consideration of the $M(T)$ and $M(H)$, separately for each Ru-content

available.

*CaMn$_{1-x}$Ru$_x$O$_3$, x = 0, 0.1:*

For $x = 0.1$, a FM order emerges with small spontaneous moment at the ground state.[11,19] Therefore, $M_{FC}(T)$ was measured in (high) magnetic field $H = 10$ kOe, as seen in Fig. 1(a); for comparison, we also display $M(T)$ of CaMnO$_3$. It is seen that $M(T)$ for CaMn$_{0.9}$Ru$_{0.1}$O$_3$ strongly decreases with increasing pressure at $T < 130$ K, nevertheless being still larger than $M(T)$ for CaMnO$_3$ which turns out to be pressure insensitive up to $P = 10.8$ kbar. Figure 1(b) shows the $H/M$ curves, plotted vs. $T$, in certainly paramagnetic (PM) range. The curves are well fitted to the Curie–Weiss law $H/M = (T - \Theta)/C$, $\Theta$ being the PM Curie temperature. For CaMn$_{0.9}$Ru$_{0.1}$O$_3$, $\Theta$ decreases with increasing pressure (remaining positive at all $P$ used) from $\Theta \approx 111.1$ K at $P = 0$ to $\Theta \approx 86.5$ K at $P = 11.3$ kbar, see Fig. 1(b). For CaMnO$_3$, at ambient pressure (not shown here) $\Theta \approx -350$ K, in agreement with the data of Ref. 22. Notice that the values of $\Theta$ for CaMnO$_3$ obtained from the fit have rather large error ± 5 K, because of small size of the studied sample, characterized by low PM magnetization, see Fig. 1(a). Nevertheless, both fitting results and closeness of the $M(T)$ dependences for $P = 0$ and $P = 10.8$ kbar, see Fig. 1(a), show that $\Theta$ is almost independent of pressure. At the same time, pressure slightly increases $T_N$ of CaMnO$_3$.[19,23]

*CaMn$_{0.8}$Ru$_{0.2}$O$_3$:*

For $x > 0.1$ the FM correlations become well developed and even small magnetic field is capable to fix an appreciable spontaneous-like magnetization. At these Ru-contents we measured $M_{FC}(T)$ and $M_{ZFC}(T)$ in $H = 100$ Oe. For CaMn$_{0.8}$Ru$_{0.2}$O$_3$, the $M(T)$ curves under various pressures are presented in Fig. 2(a). Low-temperature part of the $M_{ZFC}(T)$ curve resembles much that for an AFM. The above-defined $T_C$ decreases approximately linearly with increasing pressure from $T_C \approx 173$ K at $P = 0$ to $T_C \approx 159$ K at $P = 11.8$ kbar, thus having a pressure coefficient $dT_C/dP \approx -1.2$ K/kbar. The dependence $M(H)$ on cycling $H$ (the curve of magnetization) measured at $T = 5$ K is shown in Fig. 2(b). It is seen that (i) the coercive field $H_C$ is small; (ii) the $M(H)$ curve for large $H$ is well approximated by a small-slope straight line; (iii) the latter, extrapolated to $H = 0$, cuts the $M$-axis at some $M_0 \neq 0$, indicating the presence of FM phase. Items (ii) and (iii) points to mixed AFM-FM ground state, $M_0$ being its spontaneous magnetization. The applied pressure strongly reduces it from $M_0 \approx 0.74$ μ$_B$/f.u. at $P = 0$ to $M_0 \approx 0.54$ μ$_B$/f.u. at $P = 11.8$ kbar, see inset to Fig. 2(b). The coercive field $H_C$, estimated at ambient pressure as 0.25 kOe, slightly increases with increasing pressure.

*CaMn$_{0.7}$Ru$_{0.3}$O$_3$:*

Figure 3 displays the magnetic measurement results for CaMn$_{0.7}$Ru$_{0.3}$O$_3$. The temperature $T_C$ for this Ru-content, behaves with pressure much alike that for $x = 0.2$, see Fig. 3(a). Again, we observed a nearly linear decrease of $T_C$ with increasing pressure ($T_C \approx 199$ K at the ambient pressure) and pressure coefficient of $dT_C/dP \approx -1.3$ K/kbar, larger than that for $x = 0.2$. In addition, a cusp-like maximum at $T_N \approx 156$ K emerges on the $M_{ZFC}(T)$ curve, while being barely seen on $M_{FC}(T)$, and $T_N$ increases with increasing $P$, see Fig. 3(a). We estimated the corresponding pressure coefficient $dT_N/dP$ as equal to about 0.8 K/kbar. At highest applied pressure $P = 11.3$ kbar we have determined $T_C \approx 184$ K and $T_N \approx 165$ K. Quite similar behavior of the $M(T)$ curve was reported already for Ca$_{0.8}$Sm$_{0.2}$Mn$_{0.92}$Ru$_{0.08}$O$_3$.[13,16,24]

The $M(H)$ curves, measured at $T = 5$ K under various pressures and shown in Fig. 3(b), look qualitatively akin to those of CaMn$_{0.8}$Ru$_{0.2}$O$_3$, yet $H_C$ and $M_0$ are seemingly larger. As in the above-mentioned lower Ru-content samples, increasing pressure reduces $M_0$ and increases $H_C$. The inset to Fig. 3(b) shows the linear fit to the $M_0$ vs. $P$ data. The coercive field of CaMn$_{0.7}$Ru$_{0.3}$O$_3$ varies from $H_C \approx 0.43$ kOe at $P = 0$ to $H_C \approx 0.55$ kOe at $P = 11.3$ kbar.

*CaMn$_{1-x}$Ru$_x$O$_3$, $x = 0.4, 0.5$:*

Results of magnetic measurements under pressure for $x = 0.4$ were published recently, see Ref. 20. Figure 4 presents magnetic measurement results for $x = 0.5$ compound. For all pressures applied, the bell shape of $M_{ZFC}(T)$ is flat-topped, see Fig. 4(a) and Ref. 20. Such a top still remains on the $M_{FC}(T)$ curves for CaMn$_{0.5}$Ru$_{0.5}$O$_3$ and at ambient pressure for CaMn$_{0.6}$Ru$_{0.4}$O$_3$, see Fig. 4(a) and Ref. 20. As defined above, we extracted $T_C \approx 203$ K and $T_C \approx 185$ K for CaMn$_{0.6}$Ru$_{0.4}$O$_3$ and CaMn$_{0.5}$Ru$_{0.5}$O$_3$, respectively, at $P = 0$. The low-temperature maximum on the $M_{ZFC}(T)$ curves looks rather like a shoulder than a cusp. We figured corresponding to it $T_N \approx 115$ K and $T_N \approx 95$ K for $x = 0.4$ and 0.5, respectively. The above-mentioned shoulder-like maximum on the $M_{FC}(T)$ curve seems smeared for $x = 0.4$ and nonexistents for $x = 0.5$ at $P = 0$, but develops upon applying pressure, see Fig. 4(a). It seems that the same trend for both $T_C$ and $T_N$ (decrease and increase respectively) as observed for lower Ru-content compounds takes place for $x = 0.4$ (see Ref. 20), and in $T_C$ only for $x = 0.5$, see the inset to Fig. 4(a). The absolute values of obtained pressure coefficients of $T_C$, $-1.1$ K/kbar ($x = 0.4$) and $-0.92$ K/kbar ($x = 0.5$) are smaller than that for $x = 0.3$. On the contrary, $T_N$ of CaMn$_{0.5}$Ru$_{0.5}$O$_3$ decreases with increasing $P$, see Fig. 4(a). We estimated the pressure coefficient of $T_N$ as equal to about $-1.0$ K/kbar.

The $M(H, T = 5$ K$)$ curve for CaMn$_{0.6}$Ru$_{0.4}$O$_3$, is very similar to those for the $x = 0.2, 0.3$ compounds. The difference shows up only in an increased $H_C$, the smaller $M_0 = 0.60$ $\mu_B$/f.u. at $P = 0$, and strongly decreased pressure coefficient $-dM_0/dP \approx 0.006$ $\mu_B$/f.u./kbar, see Ref. 20. As a whole, the pressure has small effect on hysteretic loop of CaMn$_{0.6}$Ru$_{0.4}$O$_3$. However, the $M(H, T = 5$ K$)$ curve for CaMn$_{0.5}$Ru$_{0.5}$O$_3$, shown in Fig. 4(b), displays quite different trends vs. pressure. Namely, $M_0$ (smaller than the value for the case of $x = 0.4$) increases slightly, while $H_C$ decreases appreciably, see the linear fit in inset to Fig. 4(b). The hysteretic loop of CaMn$_{0.5}$Ru$_{0.5}$O$_3$ is seemingly wider than that of any other composition studied. In addition, we measured the magnetization curves at ambient pressure and elevated temperatures $T = 80$ and 130 K, see Fig. 4(c). Surprisingly, $M_0$ extracted as defined above, proves non-monotonous dependence vs. $T$, i.e. satisfies the relation $M_0(T = 5$ K$) \sim M_0(T = 130$ K$) < M_0(T = 80$ K$)$. Note that, contrary to the $M(H)$ dependence at $T = 5$ K, those at $T = 80$ and 130 K look nearly saturating in the investigated field range (i.e. they display much less pronounced increase with increasing $H$).

*Electrical conductivity*:

Figure 5 shows the results of the resistivity measurements of $CaMn_{1-x}Ru_xO_3$ with $x$ = 0.1, 0.2, 0.4, and 0.5, obtained at $P = 0$ and $P \sim 10$ kbar, and in the temperature range from 77 to 270 K. As seen, applied pressure increases $\rho$ for $x$ = 0.1, 0.2 and 0.4. On the contrary, $\rho$ for $x$ = 0.5 sample decreases upon pressure, see inset. In the studied concentration range, the temperature dependence of $\rho$ is seemingly semiconductor-like for $x$ = 0.1, 0.2 and 0.5, while being quasi-metallic and the lowest one for $x = 0.4$.

### C. Phenomenological discussion of the results

The effect of pressure on the magnetic and transport properties of $CaMn_{1-x}Ru_xO_3$ ($x = 0 – 0.5$) will be discussed in the present section in conjunction with the effect of Ru doping. Generally, Ru in $CaMn_{1-x}Ru_xO_3$ may exhibit two oxidation states, $Ru^{4+}(t_{2g})^4(e_g)^0$ and $Ru^{5+}(t_{2g})^3(e_g)^0$, but it was suggested [8-10] that for $0 < x < 0.5$ the $Ru^{5+}$ ions dominate. This may be understood as the appearance of the $Mn^{3+}$ ions in the result of quasi reaction $Mn^{4+} + Ru^{4+} \rightarrow Ru^{5+} + Mn^{3+}$. However, as it will be argued further, the excess electron is smeared rather over nearest-neighbor $Mn^{4+}$ ions than trapped by one of them. Thus an appearance of $Ru^{5+}$ ions mediates DE-like FM interaction between $Mn^{4+}(t_{2g}^3 e_g^0)$ in a cluster around $Ru^{5+}$. Furthermore, short-lifetime (virtual) $Mn^{3+}$ ions may interact with $Ru^{5+}$ and $Ru^{4+}$ ions via FM interaction like SE. The composition dependence of the magnetic properties is displayed in Fig. 6. The figure presents the magnetic moment of various Ru-doped samples at $H$ = 15 kOe and $T$ = 5 K, and the rate of the change of this quantity with $x$ as well. The experimental data presented in Figs. 2 – 4 are compared with the published ones. [9, 19, 20] Figure 6(b) shows the spontaneous magnetization $M_0$ and its rate of change with pressure vs. $x$. The spontaneous magnetization onset was found to occur at a doping of $x \sim$ 0.08, [9] which may be considered as a threshold for the formation of FM clusters. The increase in $M$(15 kOe) and $M_0$ at a relatively low doping may be accompanied by a crossover from dominantly AFM ($\Theta < 0$) to FM interactions ($\Theta > 0$) at $x$ = 0.08. [10] It appears, that the magnetization increases most steeply with increasing Ru doping in the vicinity $x \approx 0.1$ [see d$M$/d$x$ in Fig. 6(a)], then approaches a maximum at $x \approx 0.3$, followed by a monotonous linear-like diminution with further doping. The evolution of the magnetic and transport properties was previously explained by the above-mentioned quasi reaction between $Mn^{4+}$ and $Ru^{4+}$, yielding the electronic formula $Ca(Mn^{4+})_{1-2x}(Mn^{3+})_x(Ru^{5+})_xO_3$.[9] Electron magnetic resonance study has shown that at $x = 0.4$ low-temperature magnetic state is essentially inhomogeneous, comprising AFM and several magnetically non-equivalent FM phases.[11,25] At $0.5 \leq x \leq 1$, $Ru^{4+}$ ions appear and form new electronic configuration $Ca(Mn^{3+})_{1-x}(Ru^{5+})_{1-x}(Ru^{4+})_{2x-1}O_3$.[9] It is well known that in various hole-doped manganites the DE is maximal at the $Mn^{3+}$ to $Mn^{4+}$ concentration ratio of 2.3.[1,2] Applying this thumb rule to $CaMn_{1-x}Ru_xO_3$ yields $x \sim 0.4$ as a content of Ru optimal for DE like interactions and quasi metallic conductivity. As was suggested,[9] above $x = 0.4$ DE decreases while SE between emergent fixed-valence Mn and Ru ions increases gradually with increasing $x$.

Our measurements have shown that applied pressure suppresses FM phase in $CaMn_{1-x}Ru_xO_3$ samples with $x \leq 0.4$. This effect has been explained by pressure-induced electronic valence transition.[19,20] Due to comparable redox potential of $Mn^{3+} \leftrightarrow Mn^{4+}$ (1.02 eV) and $Ru^{4+} \leftrightarrow Ru^{5+}$ (1.07 eV) pairs,[26] the valence fluctuations: $Mn^{3+} + Ru^{5+} \leftrightarrow Mn^{4+} + Ru^{4+}$ are plausible. Taking into account known ionic radii – 0.65 Å ($Mn^{3+}$), 0.52 Å ($Mn^{4+}$), 0.56 Å ($Ru^{5+}$) and 0.62 Å ($Ru^{4+}$) – these fluctuations are concomitant with the volume fluctuation $\Delta V = \Delta V_{Mn} + \Delta V_{Ru} = \pm 0.3$ Å$^3$.[19] Thereby, pressure may block the fluctuations, returning the Mn and Ru valences to + 4, which leads to the suppression of DE. Such a vision is supported by decrease of resistivity with the increase of $x$ and an increase of resistivity under applied pressure, observed at $0 < x < 0.4$ (see Ref. 19 and Fig. 5). Similar scenario was already proposed for $Ba_2PrRu_{0.8}Ir_{0.2}O_6$,[27] in which the transitions $Pr^{3+} \rightarrow Pr^{4+}$ and $Ru^{5+} \rightarrow Ru^{4+}$ at $P \approx 5$ kbar are accompanied by a first-order structural transition.

Though we have recently claimed a FM phase collapse in $CaMn_{0.9}Ru_{0.1}O_3$ under pressure as a manifestation of the valence transition,[19] additional discussion of this effect is needed. In fact, the suppression of the FM phase volume in $CaMn_{0.9}Ru_{0.1}O_3$ under pressure is also accompanied by diminution of FM interactions, as seen in Fig. 1(a). Yet, $\Theta$ remains positive under pressure up to about 10 kbar and is even higher than $\Theta = 39$ K at $P = 0$ for the sample with $x = 0.08$.[9] It means that under such pressures FM correlations survive in short-range clusters, in spite of the disappearance of long range FM order.[19] The fact that the magnetization of $CaMn_{0.9}Ru_{0.1}O_3$ in PM range under pressure of about 10 kbar is much higher than that of $CaMnO_3$ supports this conclusion. Figure 6(b) shows that at $x > 0.1$ absolute value of the pressure coefficient of spontaneous magnetization $-dM_0/dP$ decreases almost linearly with increasing $x$. The reduction of $-dM_0/dP$ with increasing Ru doping [Fig. 6(b)] may be explained in part by assuming that the valence transition occurs mostly at the interface between the FM and AFM phases. Actually, the growth of the FM clusters with increasing Ru-content is accompanied by the changes in their magnetic and magneto-elastic energy, and by the lattice strain. The smaller is $x$, the smaller are the FM clusters and the larger is their surface/volume ratio. For that reason, the reduction in magnetization for the samples with smallest $x = 0.1$ and 0.15 is much larger than that for samples with larger $x$, what is indeed observed.

Pressure induced suppression of the FM phase volume by about 10% at 5 K and increase of the resistivity by about 4% at 80 K were reported for $Sm_{0.2}Ca_{0.8}Mn_{0.92}Ru_{0.08}O_3$,[16,17] so it is worth to compare $CaMn_{1-x}Ru_xO_3$ and $Sm_{0.2}Ca_{0.8}Mn_{1-x}Ru_xO_3$. The Ru doping transforms both insulator systems into quasi metals with high Curie temperatures and inhomogeneous phase separated (AFM + FM) ground states. The highest $T_C$ obtained is 240 K for $Sm_{0.2}Ca_{0.8}Mn_{0.8}Ru_{0.2}O_3$[24] and 203 K for $CaMn_{0.6}Ru_{0.4}O_3$. Recent calculations of electronic states in $Nd_{0.5}Ca_{0.5}MnO_3$ doped with various transition metal (TM) atoms on Mn-site[28] have shown that for certain impurities their $d$ states have strong effect on the host $d$ band formation. Moreover, these impurities' effects destroy CO and OO, if the impurities are set in the undoped system.[28] This study verified theoretically the unique ability of Ru to induce a FM state with both large $T_C$ and $M_0$ (e.g. $T_C = 240$ K and $M_0 \sim 2.75$ $\mu_B$/f.u. at only 5% Ru-fraction of the neodymium-calcium manganite[28]). Unfortunately, the results of Ref. 28 shed minute light on the underlying mechanism for strong effects of certain Mn-site substitution TM impurities, and there is little confidence that they apply to other manganites. Earlier, [24,27] the Ru – Mn $d$ states hybridization was pointed to as a reason for bit of effects discussed and reported above. In the theoretical section below, we qualitatively develop these ideas to explain as fully as possible the drastic differences in physical properties of parent CaMnO$_3$ and CaRuO$_3$ and mixed CaMn$_{1-x}$Ru$_x$O$_3$ compounds (subsection A) and the observed phenomena (subsection B).

### III. THEORETICAL

#### A. Outline

Figure 7 shows comparative energy-level diagram for the isolated 3$d$ and 4$d$ TM-ligand complexes, assuming that $nd$ shell of M is empty. In this familiar scheme, each $nd$-level splits off into doubly degenerate $e_g$ and triply degenerate $t_{2g}$ levels. Though being widely in use, these notations from crystal-field theory are misnomers. Actually, in addition to contribution of electrostatic interaction with the ligand ions, the splitting energy $\Delta_{cr}$ contains essential covalent contributions.[29,30] The covalence effect, i.e. the hybridation (bonding) of the $nd$ and the ligands' $p$ and $s$ orbitals, is much stronger for 4$d$ TM, e.g. Ru, than for 3$d$ TM, e.g. Mn. [21] Two physical effects displayed are worth noting. First is that the centre of gravity for 3$d$-levels lies pretty much lower (counting from a vacuum level) than that for 4$d$-levels. The second one is that the parameter $\Delta_{cr}$ is much larger, due to stronger covalence, for 4$d$ TM than for 3$d$ TM. The former effect is not important for single-TM oxides, e.g. CaMnO$_3$ or CaRuO$_3$, at all, but is essential for mixed TM oxides like CaMn$_{1-x}$Ru$_x$O$_3$. The latter effect proves crucial, when comparing the manganite and ruthenate oxides and trying to explain the physical properties of CaMn$_{1-x}$Ru$_x$O$_3$ over available $x$ range. To the best of our knowledge, the relations between the parameters, more detailed than outlined above, are eventually unknown.

The complex associated with some lattice site, say **j**, contains a fixed number of $d$-electrons if there is no charge transfer to other sites. The electrons occupy states which are labeled by the orbital ν (ε,γ), where ε and γ denote $t_{2g}$ ($xy$, $xz$, $yz$) and $e_g$ ($x^2-y^2$, $3z^2 - r^2$) states, respectively, and by the spin-projection σ (↑↓) index. The one-site many-electron Hamiltonian is the sum of ones for non-interacting electrons and electron-electron (e-e) interaction. It rules the population of one-electron energy levels shown in Fig. 7, and hierarchy of many-electron terms. The e-e interaction is described by onsite Coulomb repulsion ($U_{vv'}$) and Hund exchange ($I_{vv'}$) parameters. For isolated TM ion, all Hund parameters are equal, $I_{vv'} = I$, while two different Coulomb parameters, intra-orbital $U_{vv} = U$ and inter-orbital (ν ≠ ν') $U_{vv'} = U'$, emerge; these three parameters satisfy the Racah relation $U = U' + 2I$. For TM-ligand complex, three Hund ($I_{\varepsilon\varepsilon} \neq I_{\gamma\gamma} \neq I_{\varepsilon\gamma}$) and inter-orbital Coulomb ($U'_{\varepsilon\varepsilon} \neq U'_{\gamma\gamma} \neq U_{\varepsilon\gamma}$), as well as two intra-orbital Coulomb ($U_{\varepsilon\varepsilon} \neq U_{\gamma\gamma}$) parameters, exist. Even with the same number of $d$ electrons the terms may be crucially different, depending on the relation between $\Delta_{cr}$ and the e-e interaction parameters. E.g. $Mn^{3+}(O^{2-})_6$ and $Ru^{4+}(O^{2-})_6$ contain four $d$ electrons each, while $\Delta_{cr} \ll I$ for the former and $\Delta_{cr} \gg I$ for the latter. Thus their terms are different, being high-spin ($S = 2$) and low-spin ($S = 1$) configurations $(t_{2g})^3(e_g)^1$ for $MnO_6$ and $(t_{2g})^4(e_g)^0$ for $RuO_6$, respectively. For $RuO_6$ deviation from ionic limit is crucial due to strong covalence, which dramatically complicates reliable theoretical analysis. Spin-orbit couplings, important for addressing the spin-spin and spin-lattice relaxation, are considered elsewhere.[30] The effects of deformation on the $t_{2g}$ and $e_g$ energy levels can be described by five independent deformation-potential parameters.[31,32]

Because the hybrid orbitals of different TM-ligand complexes overlap, the electron transfer between different magnetic sites, say **j** ≠ **k**, occurs. The kinetic energy of transited electrons is described by celebrated hopping Hamiltonian. The hopping energy scales by the transfer integrals $t_{\mathbf{jk}}^{vv'}$ that comprise $p$-$d$ hybridization and Slater-Koster parts, expressing the transfer through intermediate virtually excited filled $p$-states and due to direct overlap, respectively. The Slater-Koster part is negligible for 3$d$ TM, but is very crucial for 4$d$ TM oxides due to far extension of the hybrid orbitals. For nearest-neighbor sites, the indirect $p$-$d$ hybridization part is estimated by: $t_{\mathbf{jk}}^{\mu v} \propto (1/2) V_{pd\mu}^* V_{pdv} (\Delta_{pdv}^{-1} + \Delta_{pd\mu}^{-1})$,[21] where $V_{pdv}$ is the matrix element of hybridization and $\Delta_{pd\nu}$ is the charge-transfer (CT) energy gap between the O$^{2-}$ *p* and TM hybrid *dv* levels. Deformation changes the distance |**j** − **k**| between TM ions so, changes $t_{\mathbf{jk}}^{\nu\nu'}$ by striction and via the variation of $V_{pd\nu}$ and of $\Delta_{pd\nu}$. Only speculations with regard to the pressure effects, based on specific experimental data are available at present in the literature.[23,33]

*End composition oxides*

When assembling identical TM-ligand complexes into a periodic lattice and disregarding e-e interaction, the hopping smears the levels, shown in Fig. 7, into the bands, of which widths are ∝ $|t_{\mathbf{jk}}^{\nu\nu'}|$ at a typical distance between magnetic ions. In TM oxides (unlike transition metals) $\Delta_{cr}$ > $|t_{\mathbf{jk}}^{\nu\nu'}|$ these virtual $e_g$ and $t_{2g}$ bands do not overlap. If no relations of $|t_{\mathbf{jk}}^{\nu\nu'}|$ to *U*, *U*′ and *I* are presupposed, the oxide is described by generalized Hubbard Hamiltonian, i.e. the sum of the one-electron part, including hopping, and onsite e-e interaction. Since the oxygen bonds are saturated in isolated TM-ligand complexes, in their periodic assemblies, like CaMnO$_3$ and CaRuO$_3$, the O$^{2-}$ derived *p* band is filled. In CaMnO$_3$ the *dγ* band is empty and *dε* is half-filled. Goodenough [29] discussed criteria for localized vs. collective electrons and concluded that, CaMnO$_3$ is an insulator, where electrons in the *dε* band remain localized, in agreement with experiment. They form the localized spins of Mn$^{4+}$ ions coupled via Heisenberg AFM exchange, which comprises Anderson-Hasegawa SE and Goodenough-Kanamori semi-covalent (SC) exchange contributions.[30] Lightly conventionally doped [33,34] CaMnO$_3$ becomes low resistive that hints at a small intrinsic dielectric band gap in this material. Thumb rule consideration [33] and band structure simulation [35] showed that the gap is of CT nature for CaMnO$_3$. In the band diagram of CaMnO$_3$ thus revisited [33] the *dε* band merges into the *p* band. Computation reported the CT gap to be ≈ 0.4 eV,[35] though one might not rely too much upon the number; small CT gap results in increased importance of the SC exchange.[33]

On the opposite composition side, CaRuO$_3$ is non-Fermi-liquid metallic[36] with no long range magnetic order at ground state, but with notable short-range magnetic correlations. [37] Thus, the localized-ions picture of CaRuO$_3$, where Ru resides at the fixed valence 4+ state ($S = 1$), is not adequate. Strong ligand-field factor sets the $d\varepsilon$ band to be the only relevant for the physics of CaRuO$_3$, but due to the Slater-Koster overlap this band is much broader than the like band in CaMnO$_3$. As in the case of SrRuO$_3$,[29] the $4d\varepsilon$ band delocalizes and provides metallic ground state. Concerning magnetism, these similar materials are drastically different – SrRuO$_3$ is a FM metal[29] displaying Fermi-liquid behavior at low temperatures [36] (experimental evidence of fair importance of the second coordination). Low metallic conductivity indicates that the $d\varepsilon$ band width is slightly greater than $U$. Thus the theory of (Ca,Sr)MnO$_3$ should inevitably operate generalized Hubbard Hamiltonian truncated to the $d\varepsilon$ band. This message establishes similarity between the itinerant-electron magnetism in AMnO$_3$ and in transition metals. By this analogy, one may suggest that there emerges a spin-correlated PM state if the $d\varepsilon$ band width is greater than $I$ (CaRuO$_3$), and a FM state for an opposite relation (SrRuO$_3$). First-principle theory of the ruthenates remains great challenge for theorists.

*Mixed oxides – CaMn$_{1-x}$Ru$_x$O$_3$*

When substituting Mn in CaMnO$_3$ by Ru, five Mn related $d$ states become extracted, the Ru $e_g$ and $t_{2g}$ states added, and all the $d$-states then renormalize due to hopping between Ru and Mn. Emergent by such hybridization $e_{g'}$ and $t_{2g'}$ bonding states deviate from the CaMnO$_3$ $d\gamma$ and $d\varepsilon$ bands, in respect, proportionally to the Ru portion while the anti-bonding $e_{g*}$ and $t_{2g*}$ ones split off well above the $e_g$ and $t_{2g}$ levels of the RuO$_6$ complex even at one Ru present. As seen from Fig. 7, the impurity $e_{g*}$ levels are likely irrelevant, while the $t_{2g*}$ ones should lie close to the $d\gamma$ band bottom. The electron filling changes the picture so far considered. The $t_{2g'}$ states remain filled like the Mn core $t_{2g}$ states; see Fig. 8(a), but filling of the other relevant levels becomes equivocal. From the charge neutrality viewpoint it may seem favorable to have the $(t_{2g*})^4$ term, i.e. stable Ru$^{4+}$ state, and the $d\gamma$ band empty. However, due to the e-e interaction, the level of fourth $t_{2g*}$ electron lies likely much above the $d\gamma$ band bottom at which electron gains energy of Hund exchange with the Mn core. Such a situation is depicted in Fig. 8(a), where the level of fourth $t_{2g*}$ electron is presented as narrow $Ru^{5+}/Ru^{4+}$ band, which broadens with increasing $x$. When putting the fourth $d$ electron of $Ru^{4+}$ at the $d\gamma$ band bottom one creates $Ru^{5+}$, while $Mn^{4+}$ ions remain the background. The net charge $+e$ pulls the electron back to $Ru^{5+}$ but the gain in exchange energy forces it to be near $Mn^{4+}$. Thus an impurity state emerges, where electron is smeared over nearest $Mn^{4+}$ ions, for which we use the same notation $e_{g'}$ as for the bonding states discussed above. We mark the impurity $e_{g'}$ states in Fig. 8(a) by a band adjacent to the $d\gamma$ band bottom, having in mind finite $x$. Seemingly, the $d$-electron configuration is $(t_{2g*})^3(e_{g'})^1$ as it has the lowest energy. In other words, there is a $Ru^{5+}$ ion embedded into the $Mn^{4+}$ host plus excess electron on the orbital $e_{g'}$ hopping over the nearest $Mn^{4+}$ ions, but *by no means* a stable $Mn^{3+}$ ion. But why no $Mn^{4+}$ ion captures this electron by JT effect and transform to $Mn^{3+}$ one? The reason is the hybridization of $Ru(t_{2g})$ and $Mn(e_g)$ states, not forbidden by the symmetry. The appreciable amplitudes $t^{\varepsilon\gamma}_{\mathbf{j}(Mn)\mathbf{k}(Ru)}$ and closeness of $e_{g'}$ and $t_{2g*}$ levels strongly enhance this hybridization, due to high covalence of Ru, what is unusual for parent oxides in $CaMn_{1-x}Ru_xO_3$. Supposedly high admixture of the Ru ($t_{2g}$) states in the final $e_{g'}$ states breaks $e_g$ symmetry that blocks the JT effect. Figure 8, which part was discussed above, presents one-electron energy diagrams for $CaMn_{1-x}Ru_xO_3$. Dashed lines on both parts show the levels of the $Mn^{4+}$ and $Ru^{5+}$ three core electrons, forming spin $S = 3/2$ localized on these ions. For small to moderate $x$, see Fig. 8(a), the $d\gamma$ and further $e_{g'}$ band narrows and the $Ru^{5+}/Ru^{4+}$ $t_{2g*}$ one broadens, slower than the impurity $e_{g'}$ band broadens. For larger $x$, see Fig. 8(b), the extended- and localized-states of $e_{g'}$ bands confluence, and their merger becomes narrower than the bare $d\gamma$ band was. Due to increasing overlap between $pd$ orbitals of different $RuO_6$ complexes, instead of the separate $Ru^{5+}/Ru^{4+}$ $t_{2g*}$ band and $t_{2g*}$ core levels, 2/3 filled strongly correlated band (the prototype of the $d\varepsilon$ band in $CaRuO_3$) emerges, which comprises states of mobile electrons and mixed-valence Ru ions.
15

Consider now the magnetism issue. At small to moderate $x$, see Fig. 8(a), $CaMn_{1-x}Ru_xO_3$ contains a random mixture of $Mn^{4+}$ and $Ru^{5+}$ ions, of proportion content $1 - x$ and $x$, respectively, having unique spin $S = 3/2$. According to theoretical predictions,[21] these ions are all AFM coupled. In addition, there are excess electrons interacting via Hund exchange with central $Ru^{5+}$ ions and surrounding them host $Mn^{4+}$ ions, while the relevant Hund-exchange parameters lie between large $I_{\varepsilon\gamma}(Mn)$ and much smaller $I_{\varepsilon\varepsilon}(Ru)$. The effect of carriers doped into an AFM host is well known – within spatial extension of their wave function they force the host magnetic moments to align. If the carriers move band-like, this effect results in Zener DE, which drives to FMM ground state at a low concentration of carriers,[1] but when localized, they polarize only some spin clusters around the localization centers,[38] which we will refer to as magneto-impurity state (MIS). We[7] and another group[3,4] have used the MIS concept to describe moderately electron-doped manganites. As first deduced by de Gennes,[38] MIS consists of saturated 'core' and weakly FM correlated 'halo', which so essentially affect the magnetization processes that the curves of magnetization vs. $T$ and $H$ drastically deviate from those typical for AFM. Due to the placement of the Ru impurity, the MIS model seems well suited for $CaMn_{1-x}Ru_xO$ and concords at low and moderate ($0 < x \le 0.4$) doping with Fig. 8(a). With the prerequisites delineated above it explains well both persistence of the $Ru^{5+}$ ions and the absence of monoclinic distortions (see Introduction). The $Mn^{4+}$ ions dilution at higher Ru-content affects the energy diagram and, consequently, magnetism. It may be asked why then $CaMn_{1-x}Ru_xO$ remains essentially different magnetically from $CaRuO_3$? The answer is that due to the $Ru(t_{2g})$ – $Mn(e_g)$ hybridization the $Ru^{4+}/Ru^{5+}$ $t_{2g*}$ band is not autonomous from the $Mn^{4+}$ host, and the electrons may transit from $Ru^{4+}$ to $Ru^{5+}$ in two ways. The first one is direct as in $CaRuO_3$, and the second is indirect via $Mn^{4+}$ ions or, more precisely, via virtual excitation to the $Mn^{4+}/Mn^{3+}$ $e_{g'}$ band bottom, see Fig. 8(b). The latter clearly favors the FM correlations. Also the transitions of first type act so, since the $Ru^{4+}/Ru^{5+}$ $t_{2g*}$ band width yet remains smaller than the e-e interaction parameters. Hence the competition of FM and AFM exchange interactions between the magnetic ions occurs also in this range of $x$. These interactions are essentially random due to randomness of the Ru distribution, which shows up via strong spin-glass-like effects.

### B. Theoretical discussion of experimental results

Let us now discuss in view of suggested model the experimental data obtained. For the smallest at our disposal $x = 0.1$, the band diagram is likely that as shown in Fig. 8(a) with $e_{g'}$ being isolated impurity levels. In this case, nearly isolated MIS's emerge; to fix detectable magnetization strong magnetic field is required, and hence one can hardy define such parameters as $T_C$ and $T_N$ from the magnetization vs. $T$ curve, see Fig. 1(a). Yet, a non-zero magnetization appears at ground state, likely due to the de Gennes 'halos'. The Curie-Weiss law with positive $\Theta$ fits well the data at PM range; see Fig. 1(a). The pressure acts twofold; firstly, it increases the $Mn^{4+}/Mn^{3+}$ $d\gamma$ band width via increasing the $pd$ hybridization in $MnO_6$ complexes and secondly, it decreases $|t^{\varepsilon\gamma}_{jk}|$ (since the latter should zeroing at $\mathbf{j} = \mathbf{k}$). Both effects act towards nearing (finally merging) the $e_{g'}$ level and $Mn^{4+}/Mn^{3+}$ $d\gamma$ band, as well as decreasing (finally zeroing) $I_{e_{g'}\varepsilon(Ru)}$. This drives back to $Ru^{4+}$ and results in the decrease of $\Theta$, see Fig. 1(b) and discussion above. Electric conduction in this case is facilitated by electrons activated from the $e_{g'}$ levels to the $Mn^{4+}/Mn^{3+}$ $d\gamma$ band, see Fig. 8(a), and, by the electron hopping at lower temperatures (not studied). The trend of changes of band-structure with the pressure figured above should result in increasing resistivity with increasing pressure. We indeed observed the activated behavior and the increase of $\rho(T)$ under pressure, see Fig. 5.

At larger $x$ (0.2, 0.3, 0.4), the MIS gradually enter the Anderson-localization regime, where the gap between the impurity and $Mn^{4+}/Mn^{3+}$ $e_{g'}$ bands shrinks and zeroes at the end. This leads to the appearance of both large FM moment clusters and FM exchange in the host, which, in turn, results in qualitative change of the $M(T)$ curves and magnetization process as such. In particular, $T_C$ is a temperature below which the MIS saturated 'cores' order FM but the 'halos' and the host spins are disordered, while $T_N$ is one below which the 'halos' and the host spins order AFM or in a canted structure. Thus, depending on the interplay of many parameters, such mixed AFM + FM ground states as fully PS, non-homogeneously canted or a superposition of the latter with the FM-cluster phase may be realized. With the MIS model, one expects $T_C$ to increase with increasing $x$ and to decrease with increasing $P$, which we observed indeed, see Figs. 2(a), 3(a) and Ref. 20. The same trends predicted for $M_0$, show up indeed in the $M_0$ vs. $P$ data, for $x$ = 0.2, 0.3 and 0.4, see insets to Figs. 2(b), 3(b) and Ref. 20, and in the $M_0$ vs. $x$ data, for $x$ = 0.2 and 0.3. An essential admixture of the Ru ($t_{2g}$) states, with large spin-orbital effects, adopted in the present MIS model to the $e_{g'}$ states, explains the appreciable coercive field, increasing with $x$, see Figs. 2(b), 3(b), and the unusually fast EMR line-width broadening with the increase of $x$.[30]

The observed increase of $T_N$ with $P$, see Figs. 3(a) and Ref. 20, accords well with the behavior of $T_N$ in CaMnO$_3$,[33] though $T_N$ of CaMn$_{0.7}$Ru$_{0.3}$O$_3$ is higher than that of CaMn$_{0.6}$Ru$_{0.4}$O$_3$. This effect is seemingly due to Ru$^{5+}$ – Mn$^{4+}$ and Ru$^{5+}$ – Ru$^{5+}$ SE interaction, which is larger than Mn$^{4+}$ – Mn$^{4+}$ one. At some $0.3 < x < 0.4$ $M_0$ and $T_N$ start to decrease with the increase of $x$ that indicates a crossover from the energy diagram of Fig. 8(a) to that of Fig.8(b), i.e. to fluctuating Ru-valence regime. In spite of such crossover, transport over the merged Mn$^{4+}$/Mn$^{4+}$ $e_{g'}$ band should dominate the conductivity at all moderate $x$, see Fig. 8(a), as the mobility in the emerging Ru$^{4+}$/Ru$^{5+}$ $t_{2g*}$ band is small due to strong correlations. Our resistivity data, see Fig. 5, support well this conclusion as regards to the dependence on $T$ and on $P$. Note that, though the $\rho(T)$ dependences at $x$ = 0.4 are increasing with $T$, no semiconductor-metal transition vs. $x$ occurs at $0.2 < x < 0.4$. The behavior of $\rho(T)$ observed at $x$ = 0.4, see Fig. 5, is typical of moderately doped semiconductor in the impurity exhaustion region. This region becomes unrestricted in temperature from above upon attaining $x$, at which the dielectric gap shrinks to zero, though the states remain localized – this is seemingly our case, see Fig. 5.

After the crossover to the band diagram of Fig.8(b), in between $x$ = 0.4 and 0.5, the dielectric gap increases and the conductivity becomes dominated by transport in the $Ru^{4+}/Ru^{5+}$ $t_{2g*}$ band, as discussed above. The fact that the $\rho(T)$ curve remains semiconductor-like at all pressures applied, see inset to Fig. 5, means that the electronic states remain localized. Our assumption that these are the $Ru^{4+}/Ru^{5+}$ $dt_{2g*}$ states is well confirmed by the observed decrease of the resistivity under pressure – the $Ru^{4+}/Ru^{5+}$ $t_{2g*}$ band width increases under pressure due to striction, see inset to Fig. 5 and the theoretical outline above. The fact that $H_C$ at $x = 0.5$ is much larger than $H_C$ at smaller $x$, compare Fig. 4(b) with Figs. 2(b), 3(b) and Ref. 20, also confirms the dominance of the $t_{2g*}$ states. Moreover, the behavior of $T_N$ with pressure at $x = 0.5$ is opposite to that at smaller $x$, compare Fig. 4(a) with Figs. 3(a) and Ref. 20. This behavior manifests that it is the electron transfer in the $Ru^{4+}/Ru^{5+}$ $t_{2g*}$ band, which defines $T_N$, as the pressure drives the band towards that in non-magnetic $CaRuO_3$. This is plainly not so for $T_C$ at $x = 0.5$. It behaves with the pressure in the same way as at smaller $x$, compare Fig. 4(a) with Fig. 3(a). Thus, it is the electron transfer via virtual excitation to the $Mn^{4+}/Mn^{3+}$ $e_{g'}$ band bottom, which dominates $T_C$ at $x = 0.5$. As two energy parameters $|t^{\varepsilon\gamma}_{j(Ru)k(Mn)}|$ and $\Delta_{pdv}$ rule the process, the decrease of the former is seemingly weaker than the decrease of the latter.

### IV. CONCLUSIONS

In conclusion, our studies have shown that in $CaMn_{1-x}Ru_xO_3$ ($x \leq 0.4$) the applied pressure significantly reduces the FM phase volume (characterized by spontaneous magnetization $M_0$), the electric conductivity and increases AFM interactions. On the contrary, we found that both $T_N$ and $T_C$ decrease in $CaMn_{0.5}Ru_{0.5}O_3$, while the conductivity increases, with increasing pressure. We revealed that the effect of pressure on FM and AFM correlations strongly depends on $x$, namely, the pressure suppresses the FM correlations much more pronouncedly for low doped samples ($x = 0.1, 0.15$) than for moderately doped ones ($x = 0.2, 0.3, 0.4$). For $CaMn_{0.5}Ru_{0.5}O_3$, however, pressure increases, though slightly, $M_0$ at $T = 5$ K. Starting from $x = 0.1$ the pressure coefficient

– d$M_0$/d$P$ decreases linear-like with increasing $x$. On the theoretical side, we suggested the energy diagram of CaMn$_{1-x}$Ru$_x$O$_3$, which includes a crossover from low and moderate ($x$ up to 0.4) to heavy doping ($x > 0.4$). MIS derived from this diagram at low and moderate doping, account for formation of the FM clusters and non-homogeneously canted states. At heavy doping, strongly correlated $t_{2g^*}$ band, that dominates the transport and AFM interactions, follows from our consideration. At the same time, the processes, which lead to the MIS formation at small and moderate $x$, remain to dominate the FM interactions. The predictions of the model such as Ru-valence change, the trends of $T_C$, $T_N$ and $\rho(T)$ with pressure and Ru-content, agree qualitatively with experimental data.

## ACKNOWLEDGEMENT

This research was supported by the Israeli Science Foundation (grant 209/01). The authors express their gratitude to Y. Yuzhelevskii for the assistance in transport measurements.

*Corresponding author:
Vladimir Markovich.
Department of Physics, Ben-Gurion University of the Negev,
P.O. Box 653, 84105, Beer-Sheva, Israel.
Tel: +972-8-6472456; Fax: +972-8-6472903;
E-mail: markoviv@bgumail.bgu.ac.il

**FIGURE CAPTIONS**

**Fig. 1** (a) Temperature dependence of $M_{FC}$ for $CaMn_{0.9}Ru_{0.1}O_3$ at various pressures, and for $CaMnO_3$ at $P = 0$ and $P = 10.8$ kbar, in magnetic field $H = 10$ kOe; (b) $H/M$ vs. temperature for $CaMn_{0.9}Ru_{0.1}O_3$ measured in $H = 15$ kOe above 150 K at various pressures. Solid lines are results of fitting as described in the text. Inset shows pressure dependence of the PM Curie temperature $\Theta$.

**Fig. 2** (a) Temperature dependence of $M_{FC}$ and $M_{ZFC}$ for $CaMn_{0.8}Ru_{0.2}O_3$ at various pressures in magnetic field $H = 100$ Oe; (b) Hysteretic loops of $CaMn_{0.8}Ru_{0.2}O_3$ at $T = 5$ K under various pressures. Inset shows the variation with pressure of $M_0$, as defined in the text, of $CaMn_{0.8}Ru_{0.2}O_3$ at $T = 5$ K.

**Fig. 3** (a) Temperature dependence of $M_{FC}$ and $M_{ZFC}$ for $CaMn_{0.7}Ru_{0.3}O_3$ at various pressures in magnetic field $H = 100$ Oe. (b) Hysteretic loops of $CaMn_{0.7}Ru_{0.3}O_3$ at $T = 5$ K under various pressures. Inset shows the variation of $M_0$ of $CaMn_{0.7}Ru_{0.3}O_3$ at $T = 5$ K with pressure.

**Fig. 4** (a) Temperature dependence of $M_{FC}$ and $M_{ZFC}$ for $CaMn_{0.5}Ru_{0.5}O_3$ at various pressures in magnetic field $H = 100$ Oe. Inset shows the variation of $T_C$ with pressure. (b) Hysteretic loops of $CaMn_{0.5}Ru_{0.5}O_3$ at $T = 5$ K at ambient pressure and $P = 10.8$ kbar. Inset shows the variation of coercive field at $T = 5$ K under pressure. (c) Hysteretic loops of $CaMn_{0.5}Ru_{0.5}O_3$ at ambient pressure at various temperatures.

**Fig. 5** Temperature dependence of resistivity $\rho(T)$ for $CaMn_{1-x}Ru_xO_3$ ($x = 0.1, 0.2, 0.4$) samples at ambient pressure and $P \sim 10$ kbar. Inset shows $\rho(T)$ for $x = 0.5$ sample at ambient pressure and $P = 10.2$ kbar.

**Fig. 6** (a) Variation of magnetic moment $M$ at $T = 5$ K in magnetic field $H = 15$ kOe and its $x$ derivative vs. Ru content. The data are taken from Ref. 19 for $x = 0$; 0.1; 0.15, and from Refs. 8, 9 for $x = 0.06$; 0.08; 0.6, 0.7, 0.8. Dashed line is a guide for the eye. (b) Variation of $M_0(T = 5$ K$)$ at and its pressure coefficient as a function of Ru doping; the value for $x = 0.08$ sample was evaluated from the data presented in Ref. 9. Dashed line is a guide for the eye.

**Fig. 7** Diagrams of the energy levels derived from empty $nd$-orbital embedded in TM-ligand complexes. (a) Like $MnO_6$-octahedron. (b) Like $RuO_6$-octahedron.

**Fig. 8** Suggested energy levels diagrams of $CaMn_{1-x}Ru_xO_3$. (a) Small and moderate $x$; (b) $x > 0.4$.

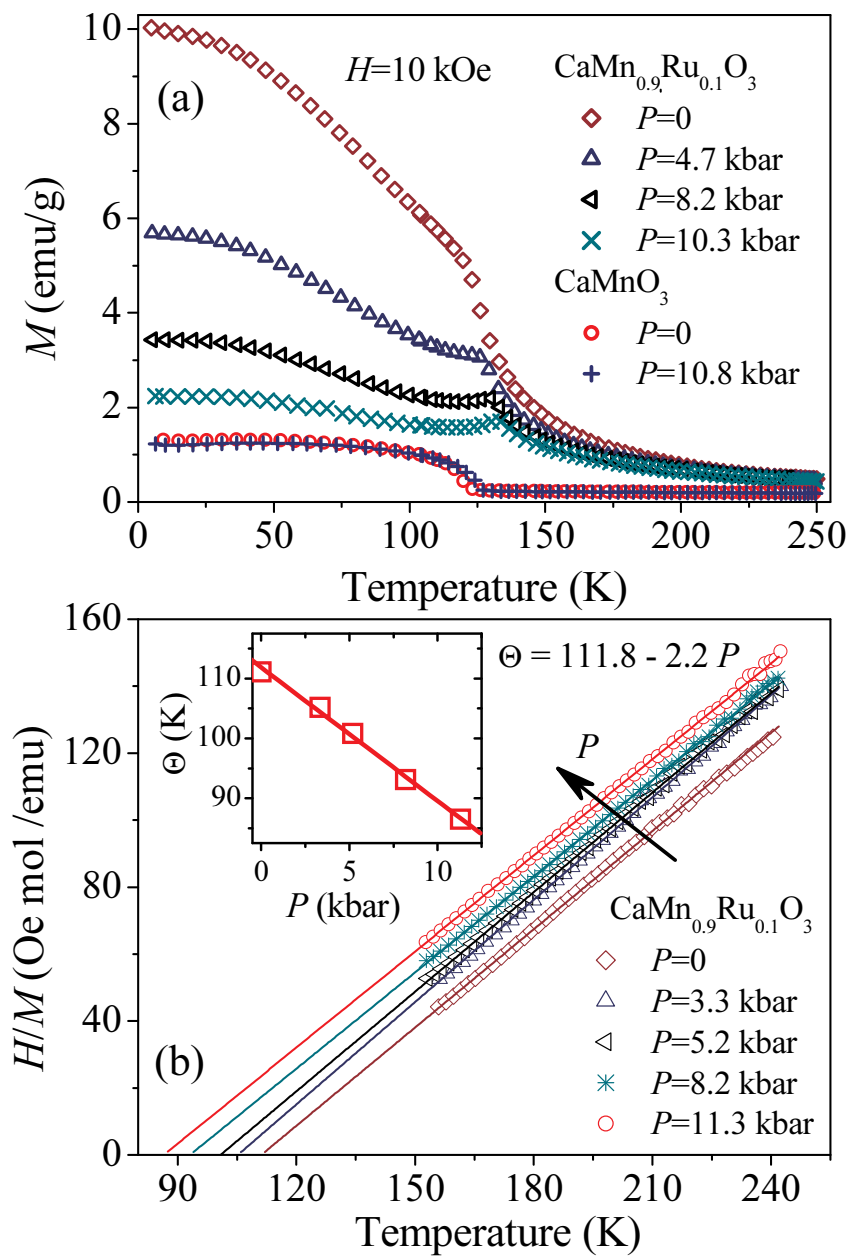

Fig. 1

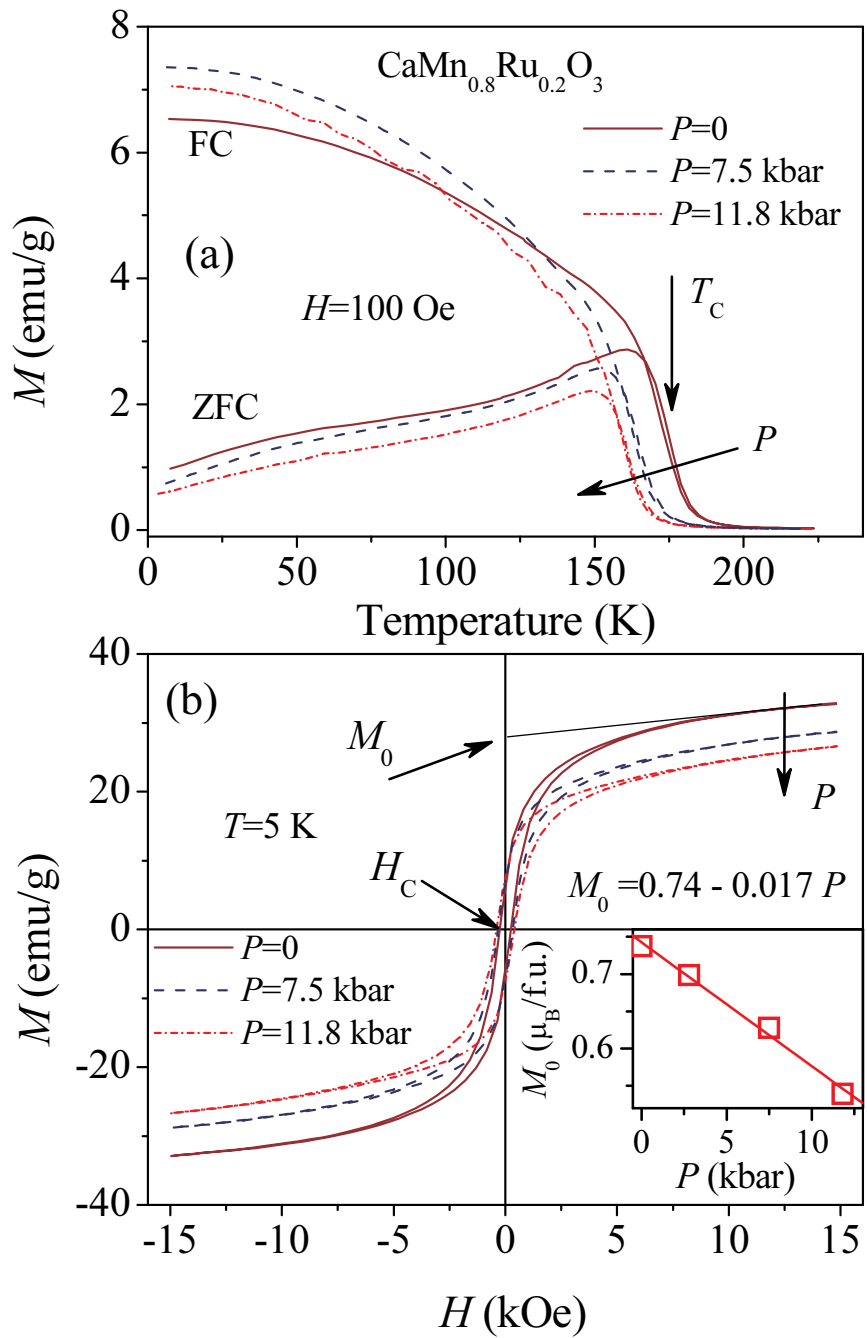

Fig. 2

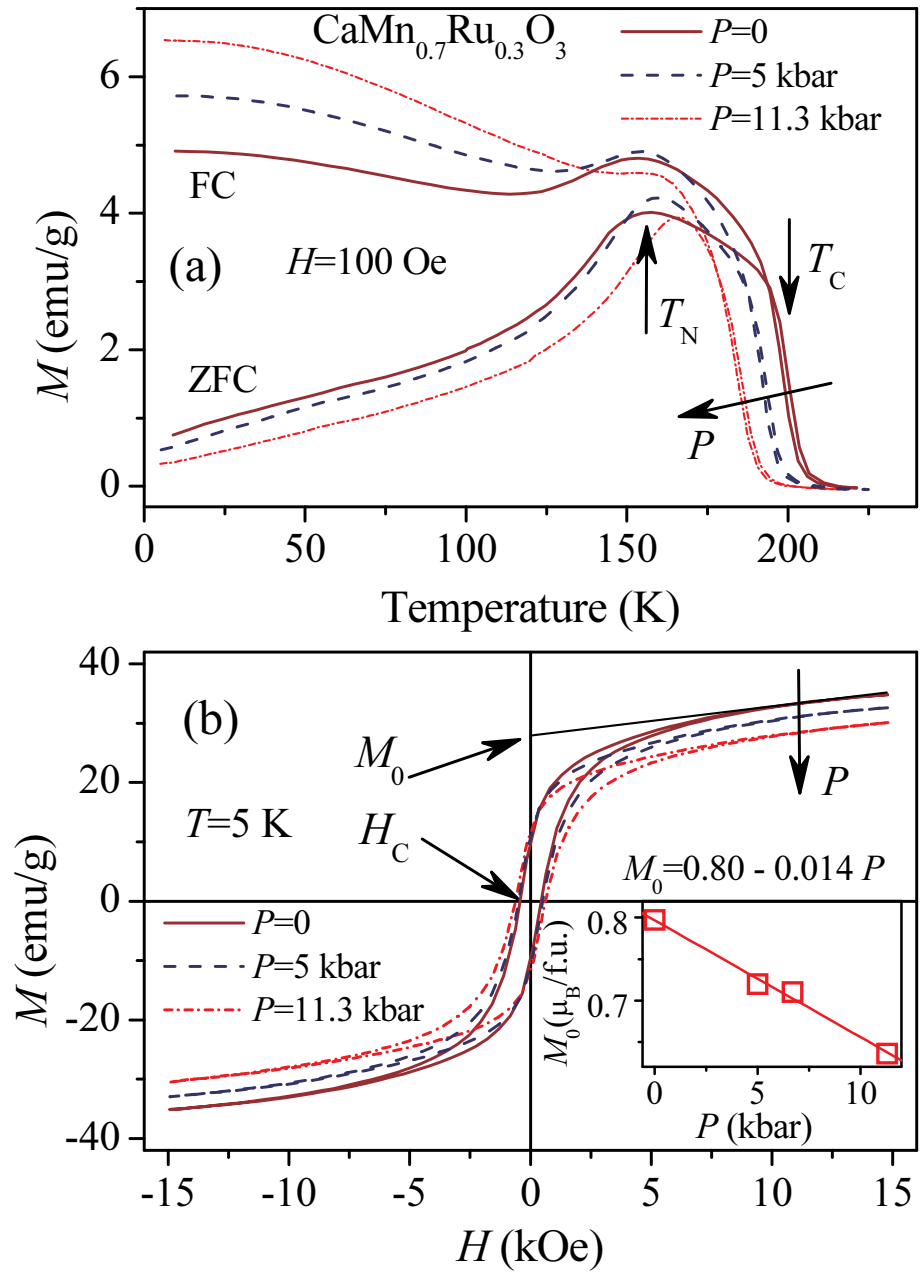

Fig. 3

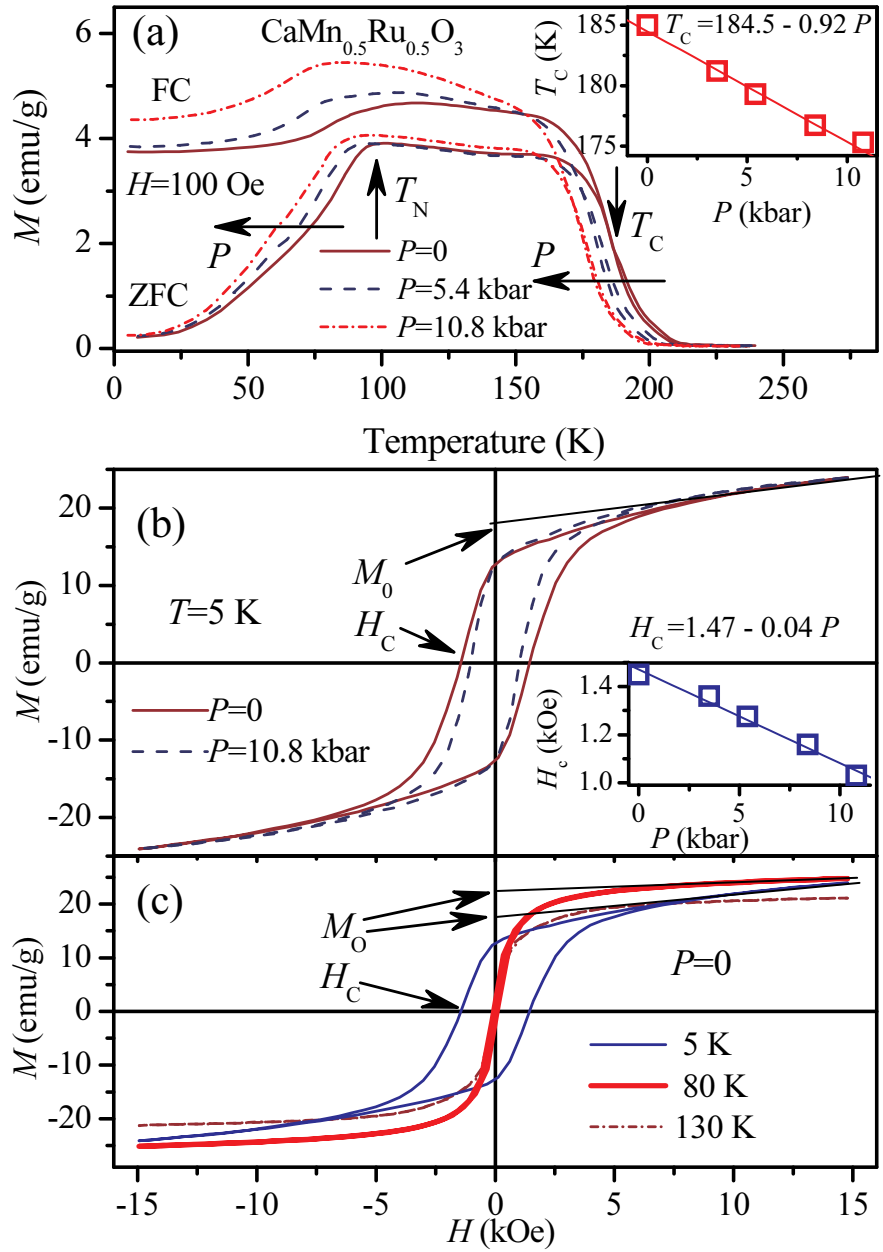

Fig. 4

Fig. 5

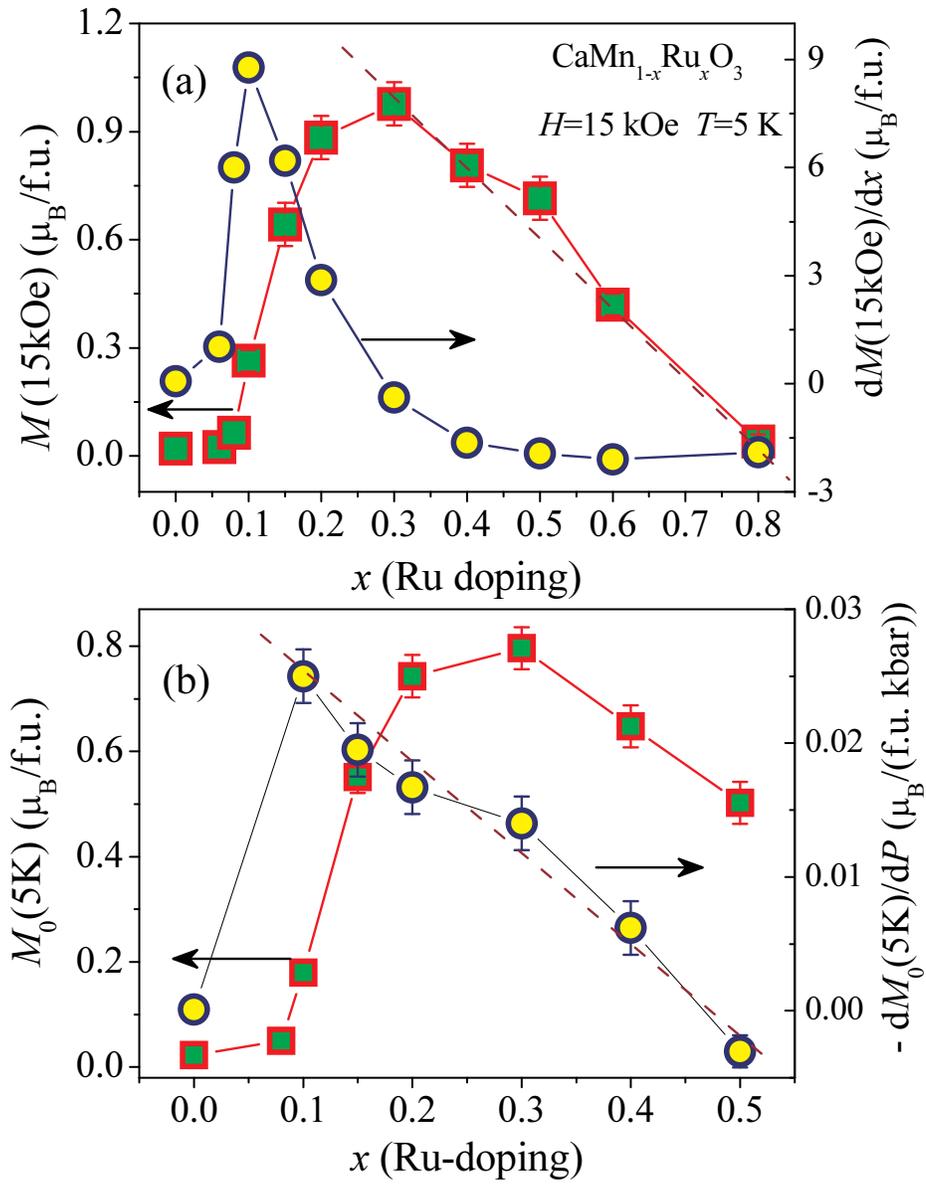

Fig. 6

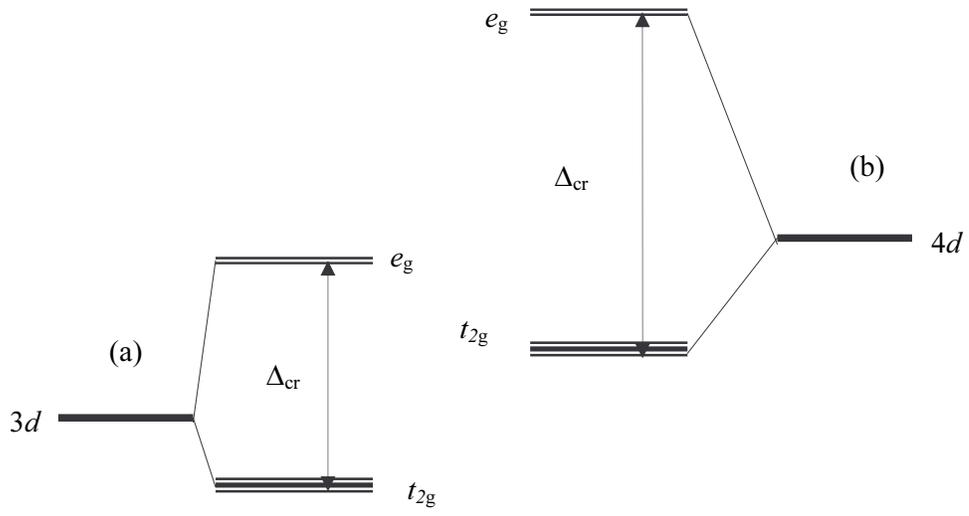

Fig. 7

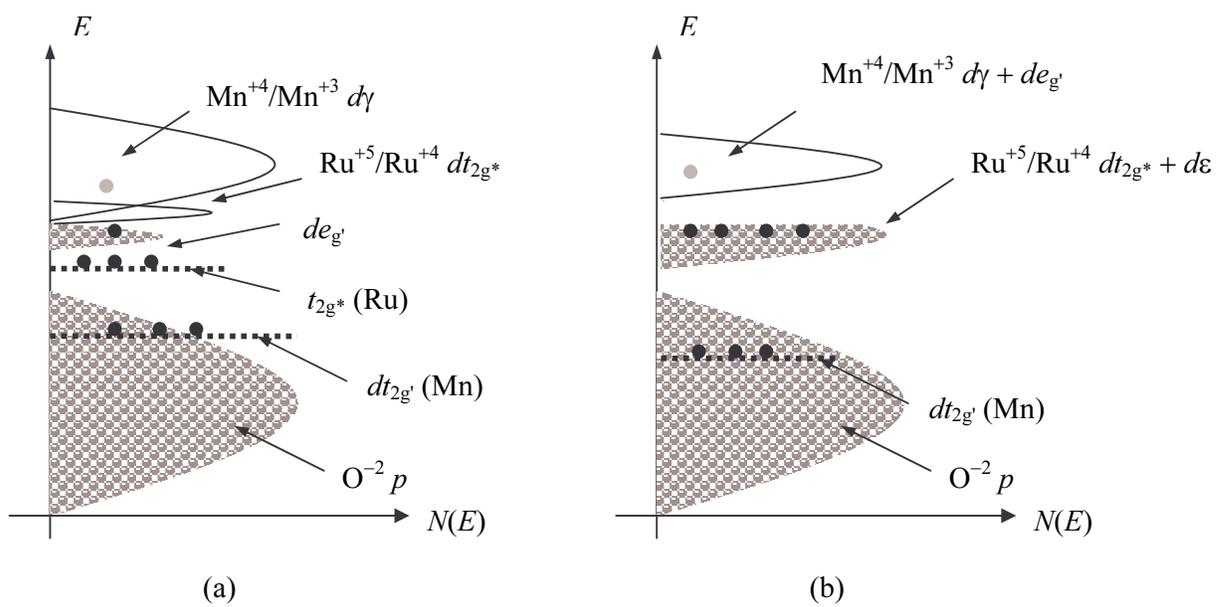

Fig. 8